\documentclass[]{aastex631}

\newcommand{\mafont}{
  \bfseries
  \color{blue}
}

\usepackage{hyperref}
\usepackage{soul}
\DeclareTextFontCommand{\ma}{\mafont}
\usepackage[normalem]{ulem}
\newcommand\maout{\bgroup\markoverwith{\textcolor{blue}{\rule[.5ex]{2pt}{1.0pt}}}\ULon}

\begin{document}

\title{Tracing Field Lines That Are Reconnecting Or Expanding Or Both}

\author{Jiong Qiu}
\affiliation{Department of Physics,
Montana State University,
Bozeman, MT 59715, USA}


\begin{abstract} 

Explosive energy release in the solar atmosphere is driven magnetically, but mechanisms triggering
the onset of the eruption remain in debate. In the case of flares and CMEs, ideal or non-ideal instabilities usually occur in the corona, but direct observations and diagnostics there are difficult to obtain. To overcome this difficulty, we analyze observational signatures in the upper chromosphere or transition region, in particular, brightenings and dimmings at the feet of coronal magnetic structures. In this paper, we examine the time evolution of spatially resolved light curves in two eruptive flares, and identify a variety of tempo-spatial sequences of brightenings and dimmings, such as dimming followed by brightening, and dimming preceded by brightening. These brightening-dimming sequences are indicative of the configuration of energy release in the form of plasma heating or bulk motion. We demonstrate the potential of using these analyses to diagnose properties of magnetic reconnection and plasma expansion in the corona during the early stage of the eruption.

\end{abstract}


\section{Introduction} \label{sec:intro}

It is well accepted that explosive energy release, in the form of solar flares and Coronal Mass Ejections (CMEs), is driven magnetically. In this process, magnetic energy is converted to kinetic energy of particles, and heat and bulk motion of plasmas \citep{Thompson2021}. CMEs open up a portion of the solar corona, along which energetic particles are released into interplanetary space and can impact the space weather. At present, routine measurements of magnetic field of the full disk are available only in the lower solar atmosphere, the photosphere. Therefore, the capability to identify solar surface signatures of open magnetic structure and to track magnetic field evolution prior to eruption will help us to explore mechanisms governing the onset of the eruption and to predict space weather.

Traditionally,  coronal holes \citep{Cranmer2009}, or regions of persistent lack of emission in soft X-ray (SXR) and extreme ultraviolet (EUV) wavelengths, are considered to map the feet of open field lines on the solar surface. The life-time of coronal holes spans many days to multiple weeks \citep[][and references therein]{Lowder2017}. The temporary, and often abrupt, opening up of the magnetic structure associated with a CME produces transient coronal holes, also called coronal dimmings \citep{Sterling1997}. A comprehensive review of observational signatures and interpretation of coronal dimmings has been provided by Veronig et al. (2024, in press). Mechanisms causing the explosive ``open-up" of the corona include the onset of an ideal instability, in which the global force balance is lost during the quasi-equilibrium evolution of the system \citep{Forbes1991, Torok2005, Kliem2006, Isenberg2007}, or the onset of a non-ideal instability, often referred to as magnetic reconnection, which abruptly changes the connectivity of field lines, and in this way removes or weakens the constraints that would maintain the force balance of a coronal structure. The ``tether-cutting" \citep{Moore2001} and ``break-out" \citep{Antiochos1999} configurations have been the most well-known reconnection geometry, which occurs either below or above the erupting structure, be it a magnetic flux rope or a sheared arcade \citep{Patsourakos20}. 

In the case of major solar eruptive events, the onset of eruption, either due to ideal or non-ideal instabilities, takes place in the corona, yet perturbations propagate along magnetic field lines to reach the lower atmosphere on Alfv{\'e}nic timescales. In this paper, we focus on dimming and brightening signatures observed at the feet or base of magnetic structures that are undergoing dynamic evolution such as magnetic reconnection or plasma expansion/eruption. As a direct consequence of magnetic reconnection, energy flux, via particle beams \citep{Fisher1985}, thermal conduction \citep{Longcope2014}, or Alfv{\'e}n waves \citep{Fletcher2008}, is transported along newly reconnected field lines to be deposited in the denser lower atmosphere, producing enhanced brightening there. On the other hand, dimming at the base of the corona is primarily an effect of plasma rarefaction, due to expansion of the overlying coronal structure such as a CME, that reduces the pressure of the overlying corona. The expansion or eruption of a coronal structure may occur before or after the onset of magnetic reconnection, therefore dimmings may be observed to either precede or follow brightenings in the lower atmosphere.

A solar eruptive event involves rather complex magnetic configuration, and different parts of the system undergo different dynamics and also interact with each other such as by reconnetion. The spatially resolved, full-disk observations, like those provided by the Solar Dynamics Observatory \citep[SDO;][]{Pesnell2012}, allows us to infer the evolution of different parts of the coronal structures during or before the eruption by examining the behavior at their feet in the lower atmosphere.  In this paper, we conduct such an experiment on two eruptive flares, using the tempo-spatial sequence of brightenings and dimmings to reconstruct the evolution
of coronal structures in the early phase of their eruption. In the following text, we describe the strategy to identify dimming signatures at the base of the corona (S\ref{sec:method}). We apply the analysis to observations of an X-class eruptive event SOL20120712 exhibiting post-eruption dimming (S\ref{sec:20120712}), and of a C-class eruptive event SOL20110621 also exhibiting pre-eruption dimming (S\ref{sec:20110621}), using these signatures to infer properties of the overlying coronal structure. We summarize what is learned from this experiment, and discuss the potential to reconstruct the evolution of overlying coronal structures using brightening and dimming signatures at their feet (S\ref{sec:summary}).

\begin{figure}    
\includegraphics[width=1.0\textwidth,clip=]{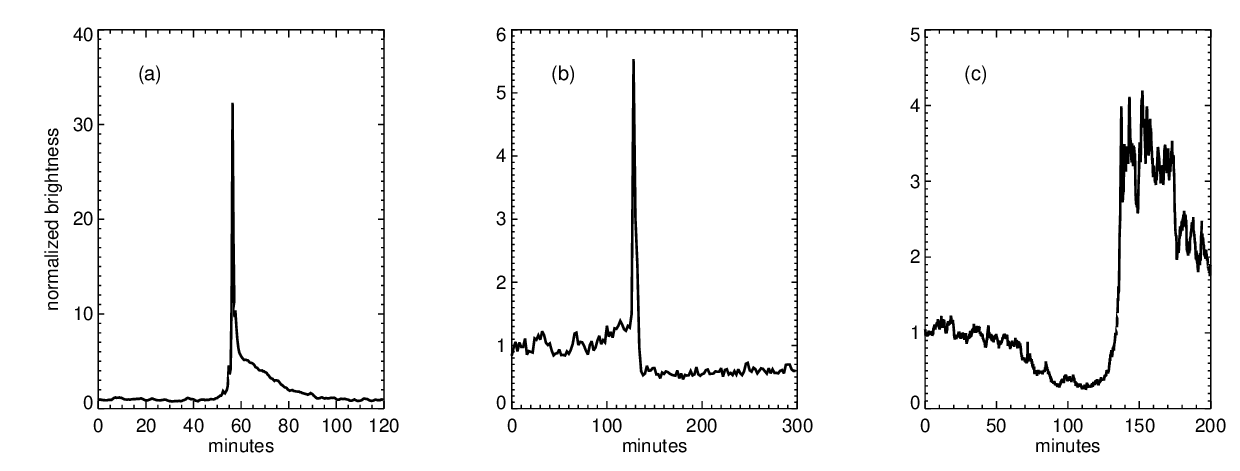}         
	\caption{Light curves of the brightness in UV 1600 or EUV 304 passbands in several pixels from various flares observed by AIA, exhibiting (a) impulsive brightening followed by prolonged brightening, (b) rapid dimming following impulsive brightening, or (c) gradual dimming before brightening.}
 \label{fig:pxlcv}  
   \end{figure}

\section{Tempo-spatial Sequence of Brightening and Dimming}\label{sec:method}

In this paper, we analyze the brightening and dimming signatures in the lower atmosphere in two eruptive events. They exhibit a variety of dimming-brightening sequence, which can be used to diagnose the dynamic evolution of overlying coronal structures in the early phase of the eruption.

\subsection{Impulsive and Prolonged Brightening}

We identify brightenings in the lower atmosphere, as these locations usually map the feet of reconnecting field lines along which energy flux is transported and deposited at the denser lower atmosphere. It has been a well adopted practice to integrate magnetic flux in these areas of the brightening in the lower atmosphere, as an estimate of the amount of reconnection flux $\psi_{rec}$ and its time derivative, the reconnection rate $\dot{\psi}_{rec}$ \citep{Fletcher2001, Asai2004, Qiu2004, Qiu2010, Kazachenko2017}. Typically observations in the optical (such as the H$_{alpha}$ line) and ultraviolet (such as the UV 1600~\AA\ passband) wavelengths are analyzed to identify brightening signatures. In this study, we analyze either the 1600\AA\ or the EUV 304~\AA\ observations from the Atmosphere Imaging Assembly \citep[AIA;][]{Lemen2012}. The contribution to the 1600~\AA\ broadband emission includes the continuum formed in the temperature minimum region and, in particular during the flare, optically-thin lines like C{\sc iv} formed at the transition -region temperature of 100,000~K \citep{Simoes2019}. The contribution to the EUV 304~\AA\ broad band is more complex, from the upper-chromosphere, transition region, and low corona, at the temperature 0.01 - 1~MK \citep{Odwyer2010}. The EUV 304~\AA\ passband is more sensitive to weak brightenings in the upper-chromosphere and transition region; on the other hand, structures in the corona, such as filaments, active region loops, and flare loops, are often observed in this passband, and are sometimes hard to be distinguished from the brightening at the feet of the coronal structures.

When magnetic reconnection occurs in the corona, energy flux along newly reconnected field lines travels to the lower atmosphere on Alfv{\'e}nic times scales, and produces a rapid or {\em impulsive brightening} there. If reconnection forms closed field lines (post-reconnection flare loops), impulsive energy release in these loops drives chromosphere evaporation that significantly raises the density (as well as the temperature) of plasmas trapped in the loops. As a result, {\em prolonged brightening} is often observed at the feet of the post-reconnection flare loops for an extended duration, of more than 10 minutes, before the brightness is attenuated to pre-flare level, reflecting the timescales of chromospheric evaporation, cooling of the heated corona, and often the continuous gradual heating \citep{Qiu2013, Qiu2016}. Therefore, light curves similar to that shown in Figure~\ref{fig:pxlcv}a, are indicative of reconnection forming closed field lines or flare loops.  On the other hand, reconnection leading to open field lines, which do not trap plasmas, would produce only impulsive brightening at its feet, as shown in Figure~\ref{fig:pxlcv}b. The pixel light curves are therefore indicative of reconnection geometry.

In this study, we identify brightening pixels in UV 1600~\AA\ passband or EUV 304~\AA\ passband, when the pixel brightness $I$ is enhanced to be more than $N$ times of their base brightness $I_0$ for more than $\tau$ minutes; $I_0$ is the average of the brightness over 20 minutes during the quiescent (pre-flare) period. $N$ and $\tau$ are empirically selected to pick out as many as possible brightening pixels in the lower atmosphere while minimizing contribution by brightenings of coronal loops. 

\subsection{Rapid and Gradual Dimming}

The majority of coronal dimming observations are made in the soft X-ray and EUV wavelengths \citep{Sterling1997, Thompson1998, Qiu2007, Mandrini2007, Temmer2017, Dissauer2018, Wang2019}. In this study, we examine signatures in the upper-chromosphere or transition region, where coronal structures undergoing dynamic evolution are anchored. We identify dimming signatures at the base of the corona primarily using observations in He~{\sc II} 304 passband on AIA, complemented with analysis of more conventional observations in the EUV passbands, including the EUV 171, 193, and 211~\AA\ passbands. The other EUV passbands are sensitive to temperatures $\ge$ 1~MK \citep{Odwyer2010}, of plasmas in the corona. As such, dimmings identified in the EUV 304 passband tend to occupy smaller areas than those identified in other bands which often include dimming signatures due to removal or re-orientation of coronal loops along the line of sight\citep{Harvey2002, Harra2007, Qiu2007, Downs2015}. In other words, detecting dimming in transition region lines helps minimize the projection effect.

We examine the time evolution of the brightness at each pixel, normalized to its base brightness. The base brightness is the mean brightness over 20-30 minutes prior to the eruption. The dimming pixels are identified if the brightness is reduced to $\le$ 80\% of the base brightness continuously for $\ge$10 minutes. The choice of the minimum dimming depth at 80\% and the minimum dimming duration of 10 minutes is mostly empirical, justified by the statistical performance of the pixel brightness. For example, the fluctuations of individual pixel's quiescent brightness is found to be about 10\%; therefore, a persistent decrease of the brightness at 20\% below the base brightness is considered to reflect genuine dimming.

We assume that these dimming pixels map the feet of magnetic structure that are expanding or erupting in the corona. Here we define ``expansion" and ``eruption" as the global or average bulk motion of the plasma in a magnetic structure, the former referring to motion at subsonic speeds and the latter at super-sonic speeds. If dimming is preceded by brightening at the same or adjacent locations, it is likely an indication that reconnection opens up overlying field lines, and the magnetic flux integrated in the brightening area provides an estimate of the amount of flux removed from above. Furthermore, the dynamic properties, such as the mean speed of the expansion of the overlying corona, can be also estimated from the evolution of the dimming depth. 

Pixel light curves in Figure~\ref{fig:pxlcv}  illustrate various dimming signatures, such as {\em rapid dimming} following impulsive brightening (panel b) indicative of reconnection opening up field lines along which plasmas rapidly expand, or {\em gradual dimming} over tens of minutes followed by brightening or rapid dimming (panel c), suggesting quasi-equilibrium expansion of the overlying coronal structure before the onset of reconnection or eruption. The tempo-spatial sequence of dimming and brightening therefore provides clues to the dynamic evolution of the corona.




\section{Reconnection-driven Post-eruption Dimming} \label{sec:20120712}

\begin{figure}    
\centerline{\includegraphics[width=1.0\textwidth,clip=]{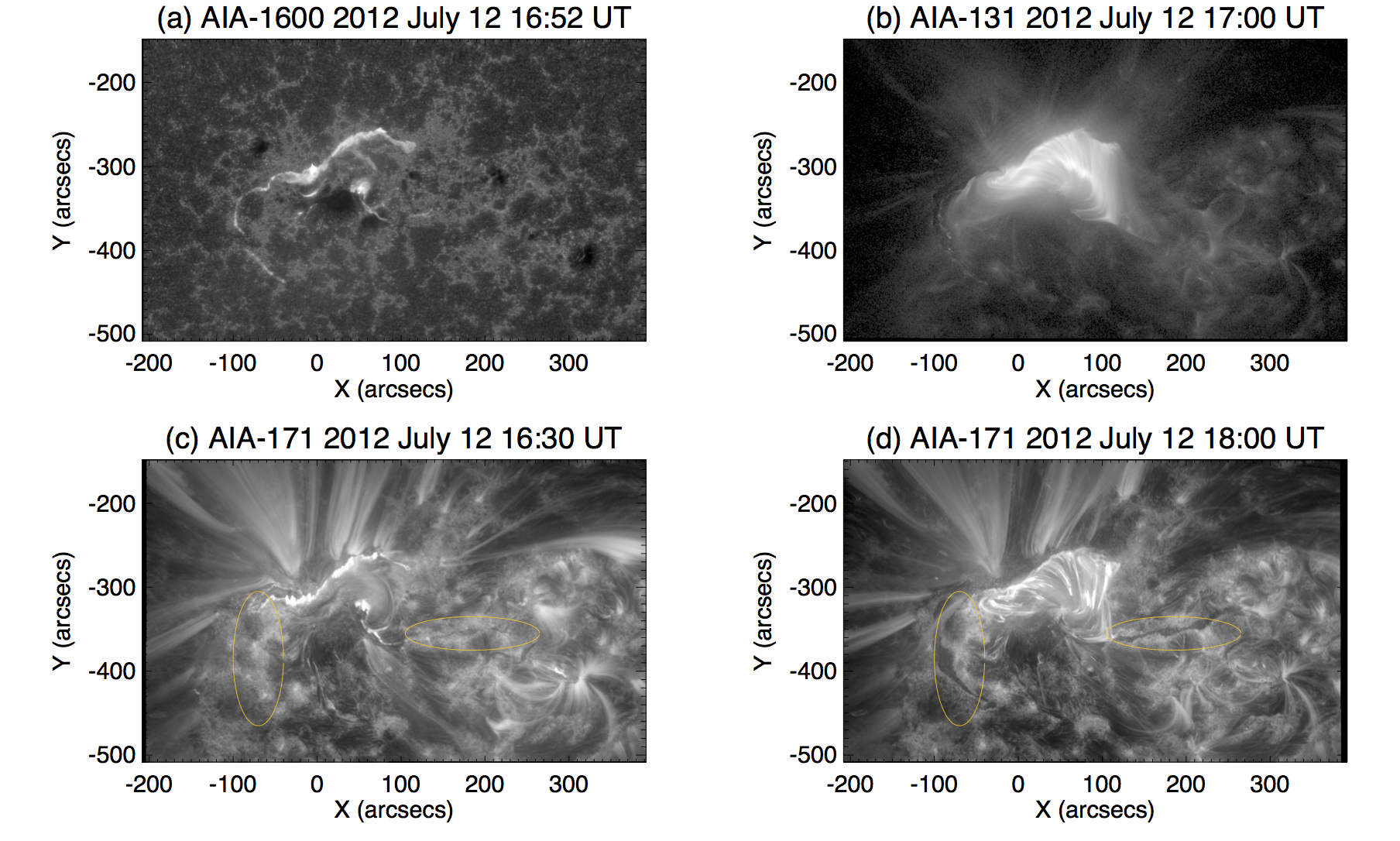}
              }         
	\caption{Overview of the SOL20120712 eruptive flare observed by AIA. Top: flare ribbons observed in UV 1600~\AA\ passband (a) and flare loops observed in EUV 131~\AA\ passband (b). Bottom: flare ribbons and loops observed in the EUV 171~\AA\ passband at two times, showing a pair of post-eruption dimming ribbons (their locations indicated by the two orange ovals) extending from the ends of the two flare ribbons. All images are co-aligned to 17:00~UT. }
 \label{fig:0712overview}  
   \end{figure}

   \begin{figure}    
    \includegraphics[width=0.50\linewidth,clip=]{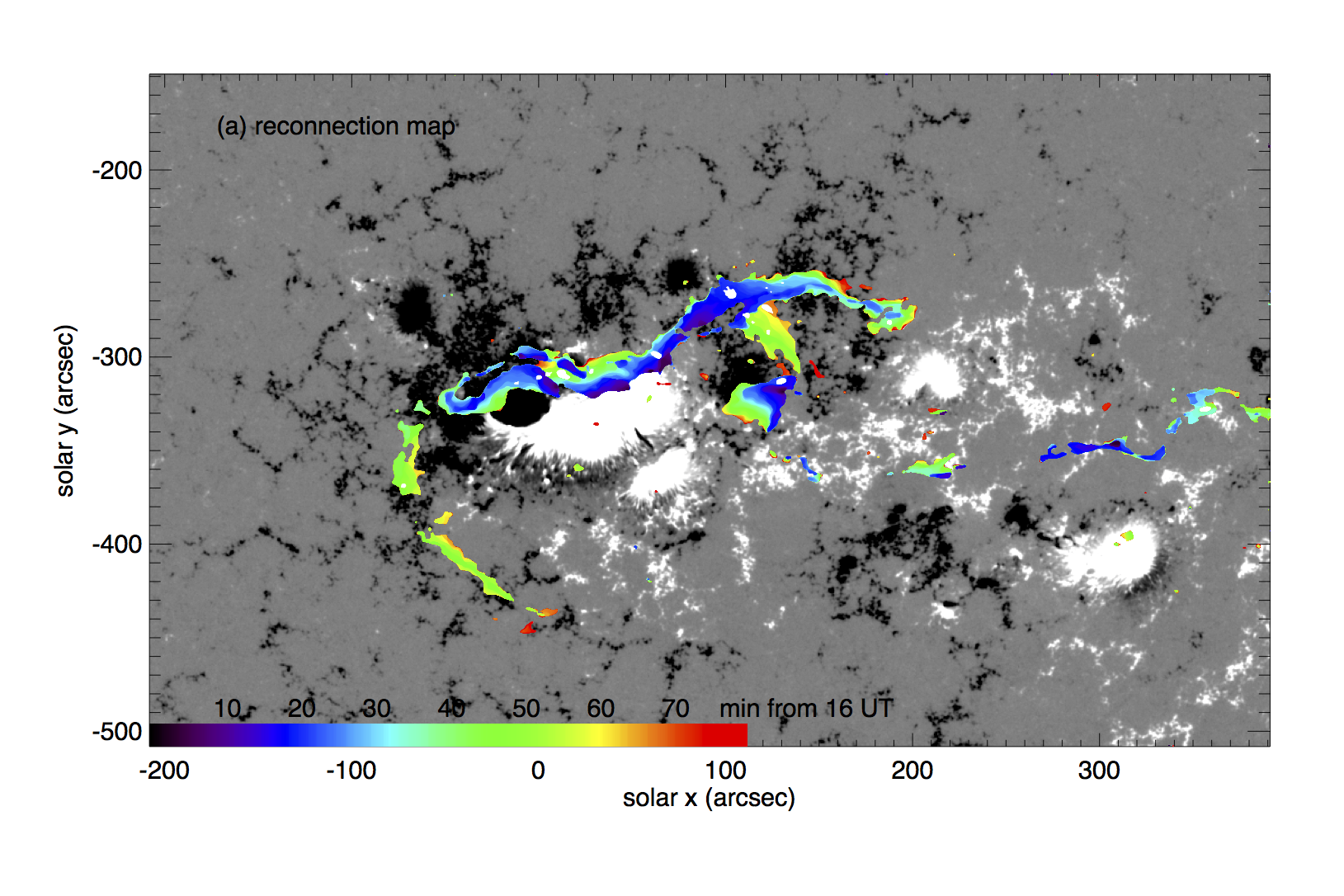}
     \includegraphics[width=0.50\linewidth,clip=]{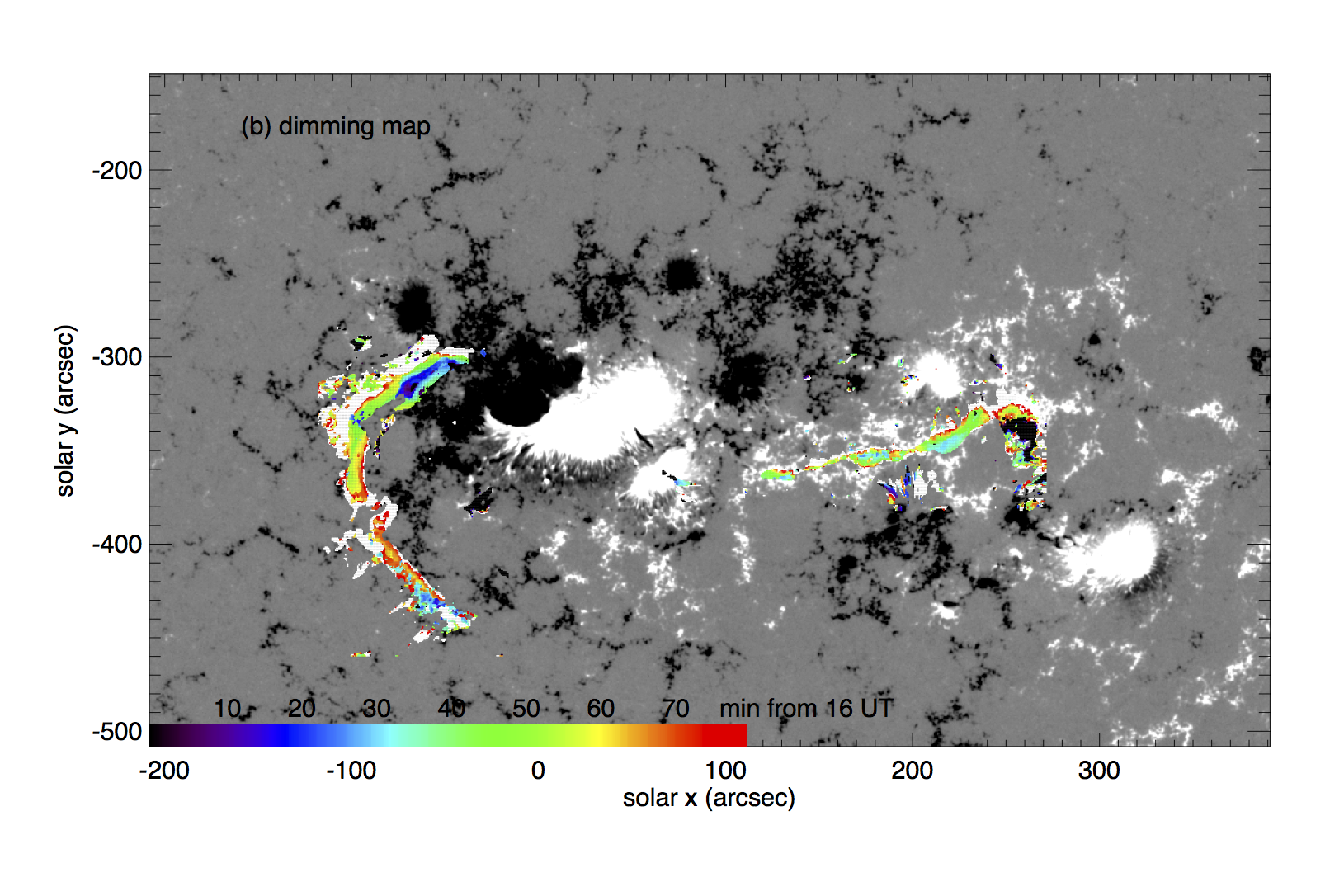}
        \includegraphics[width=0.50\linewidth,clip=]{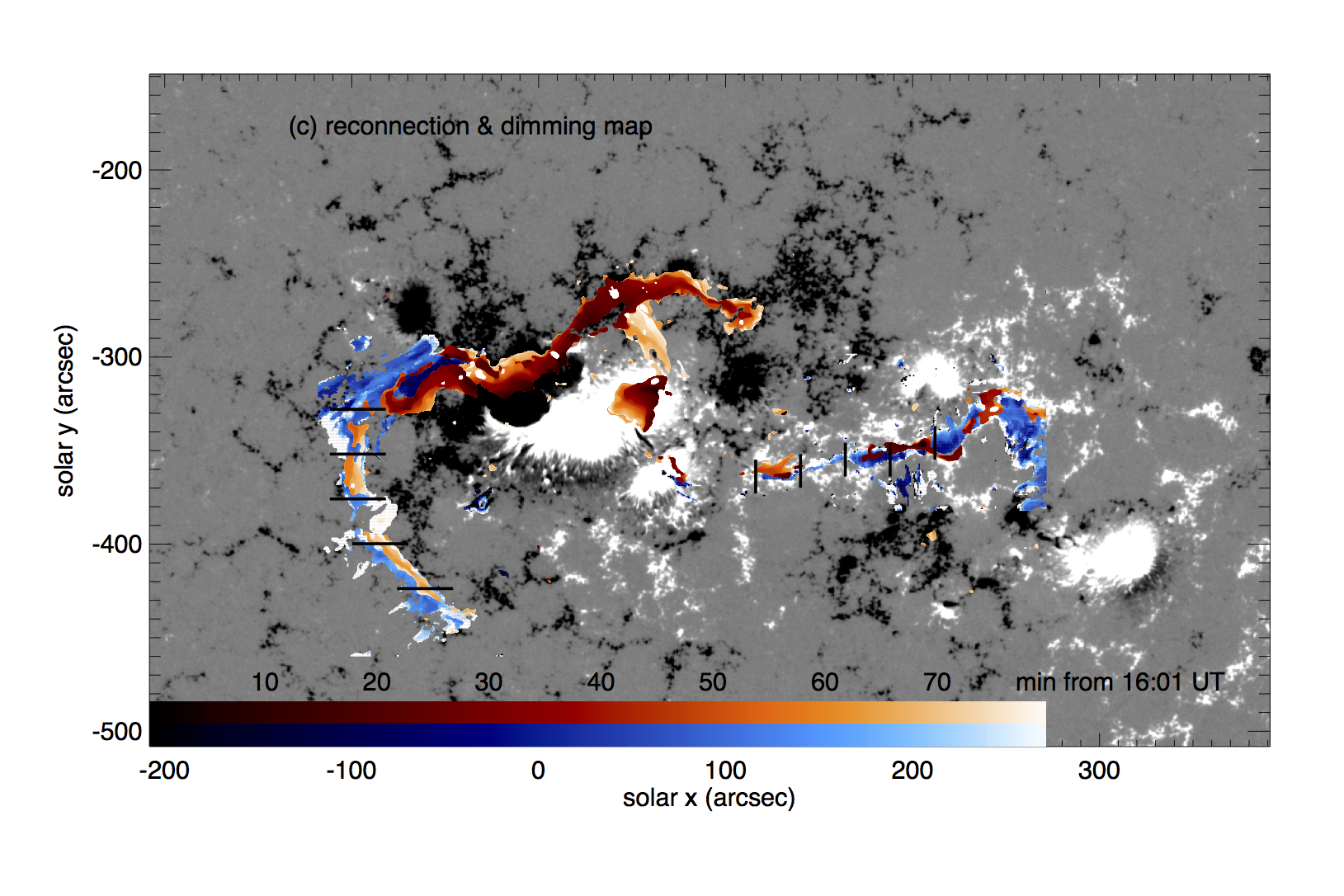}
        \includegraphics[width=0.50\linewidth,height=6.3cm]{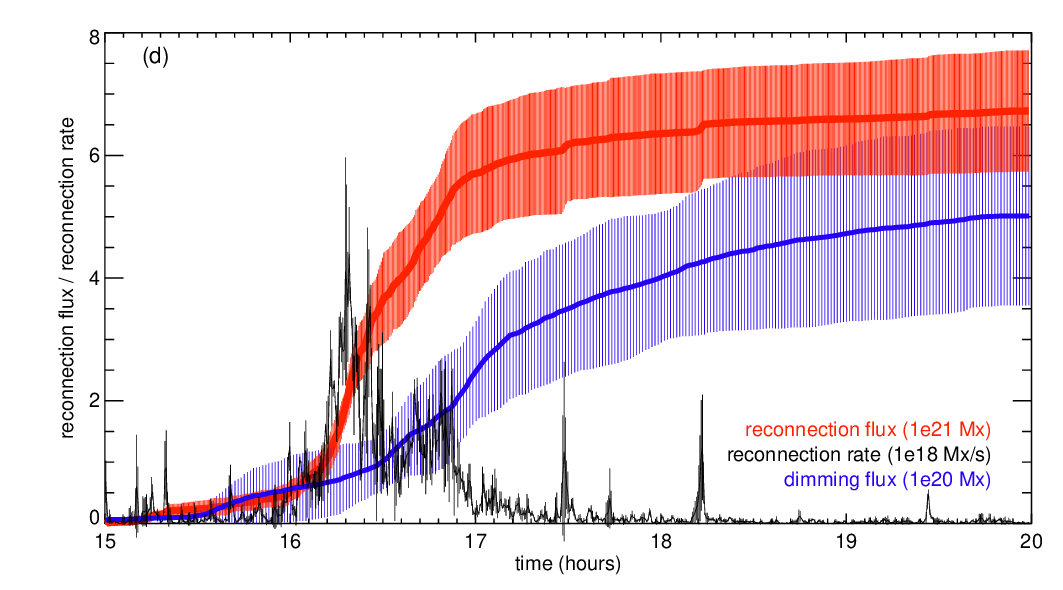}
              \caption{ Evolution of brightening and dimming. (a) Mapping of the brightening (from AIA 1600~\AA) evolution on a photospheric magnetogram (grey-scale) of the longitudinal magnetic field component $B_{los}$. (b) Mapping of the dimming (from AIA 304~\AA) evolution on the photospheric magnetogram.
              (c) Mapping of brightening (orange) and dimming (blue) on the photospheric magnetogram.
              (d) Reconnection flux $\psi_{rec}$ and reconnection rate $\dot{\psi}_{rec}$ measured from the brightening map, and the dimming flux $\psi_{dim}$ measured from the dimming map. In panels (a)-(c), the greyscale of the $B_{los}$ map is saturated at $\pm$300~G. In panels (a) and (b), the rainbow color indicates the onset time of brightening or dimming, respectively. In panel (c), the orange (blue) color scheme indicates the onset time of brightening (dimming) at a given location. Note that on the dimming ribbons, the brightening is covered by the subsequent dimming. The five horizontal bars across the dimming ribbon in the east and the five vertical bars across the dimming ribbon in the west denote the locations of the time-distance diagrams of the normalized brightness in Figure~\ref{fig:0712slit}.  }
               \label{fig:0712map}  
\end{figure}

\begin{figure}    
\includegraphics[width=0.34\textwidth,clip=]{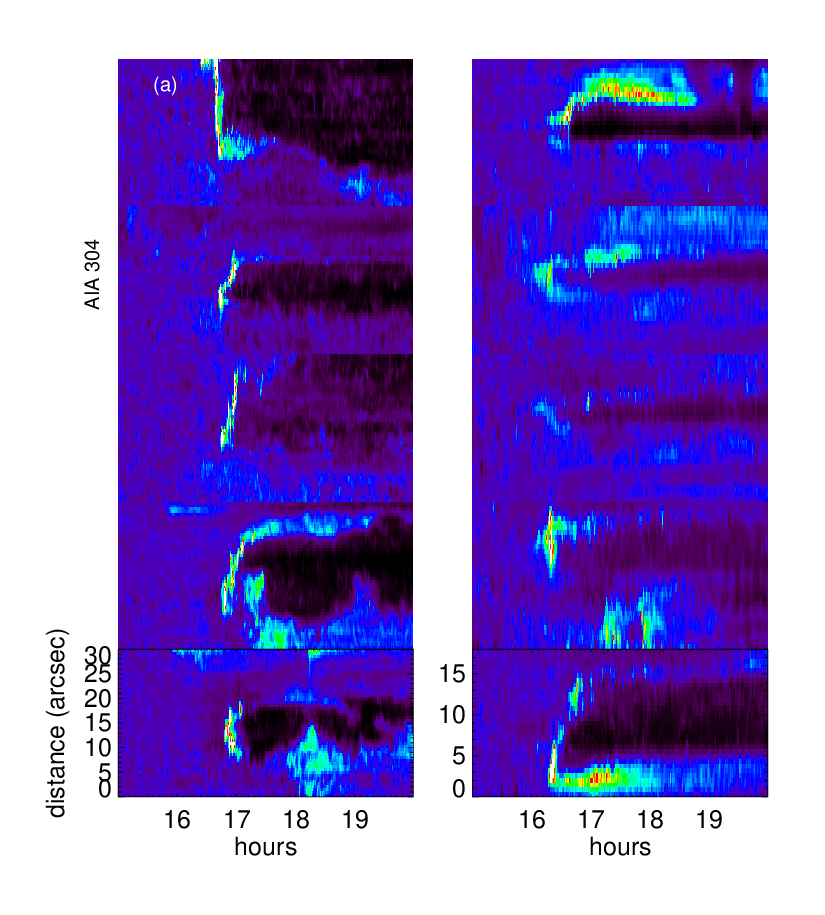}
\includegraphics[width=0.34\textwidth,clip=]{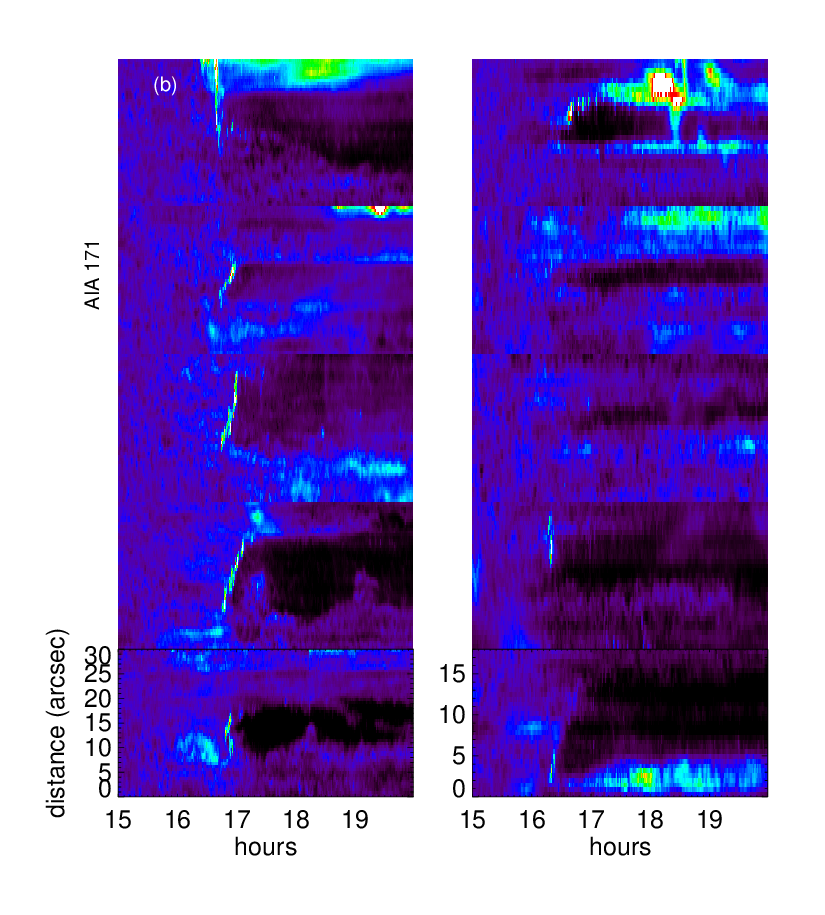}
\includegraphics[width=0.34\textwidth,clip=]{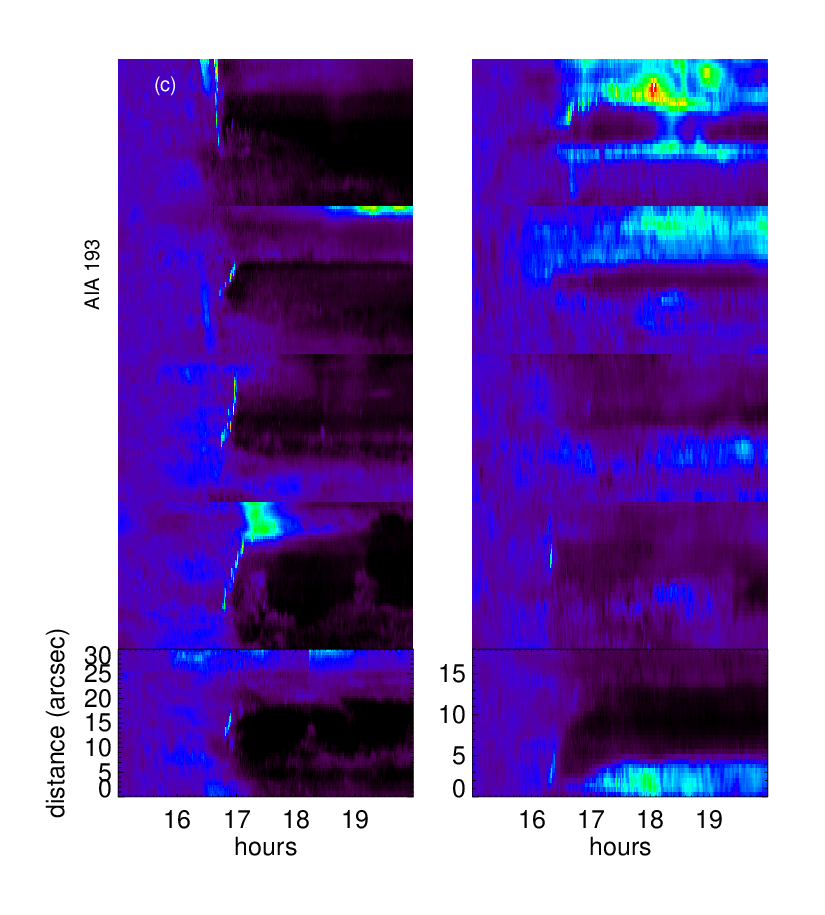}
	\caption{Time-distance diagrams of the normalized brightness along the five slits from top to bottom across the dimming ribbon in the east (left), and along the five slits from left to right across the dimming ribbon in the west (right), in 304~\AA\ (a),  171~\AA\ (b), and 193~\AA\ (c) passbands, respectively, all exhibiting rapid dimming after impulsive brightening.
 }
\label{fig:0712slit}  
   \end{figure}

\begin{figure}    
\includegraphics[width=0.98\textwidth,clip=]{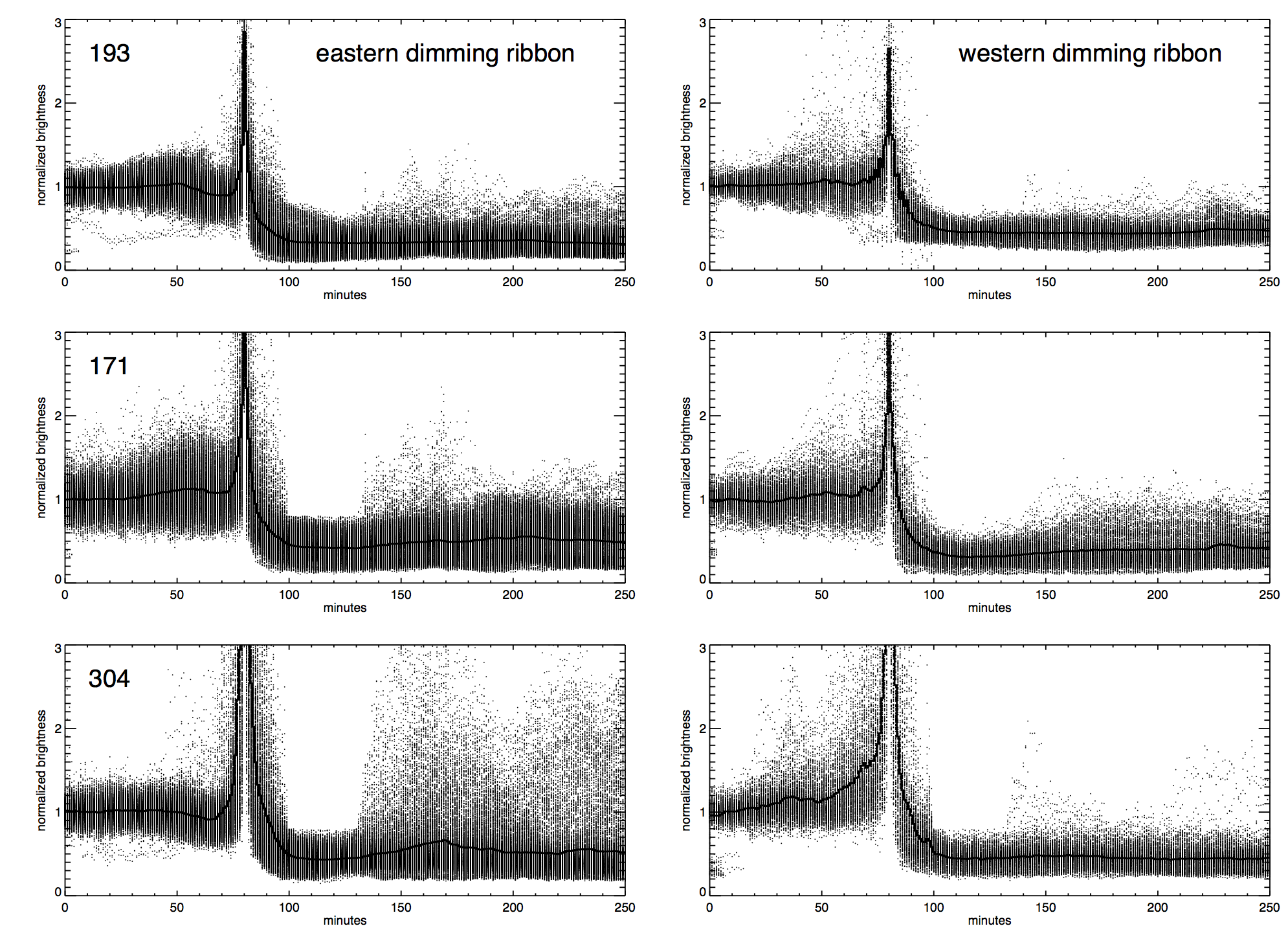}
	\caption{Epoch plots of the light curves of the normalized brightness of about 1000 pixels in the eastern dimming ribbon (left), and of about 300 pixels in the  western dimming ribbon (right), observed in EUV 193~\AA\, 171~\AA\, and 304~\AA\ passbands, respectively.
 }
\label{fig:0712epoch}  
   \end{figure}

An X-class eruptive flare occurred on 2012 July 12 in NOAA-11520. The flare is accompanied by a fast CME with its early-phase motion best captured by STEREO-A \citep{Cheng2014, Dudik2014, Zhu2020}. Figure~\ref{fig:0712overview} gives several snapshots of the event observed by AIA. The eruption occurs in the core active region forming bright two ribbons in magnetic fields of opposite polarities, outlining the feet of post-reconnection flare loops, or closed field lines produced by reconnection. We refer to the two-ribbons as {\em flare ribbons}. Shortly afterwards, another set of two long ribbons extend from the flare ribbons at the core region, which are briefly brightened initially, and then quickly become dark. We call these two ribbons that become dark {\em dimming ribbons}, and they are also located in magnetic fields of opposite polarities. 

Figure~\ref{fig:0712map}a shows, in red color, the time sequence of brightening in {\em flare ribbons} in the core region and also the brightening adjacent to the dimming ribbons away from the core region. 
This reconnection mask is derived using the UV 1600~\AA\ observations, with the method described in \citet{Qiu2007, Qiu2010}. In the same figure, the blue color shows the time sequence of the dimming, when the brightness is reduced to 80\% of the base level (see S\ref{sec:method} for identification of the dimming pixels). Dimming (and the preceding brightening) appears to spread rapidly along the dimming ribbons at an apparent speed of more than 100~km $^{-1}$, noted as signatures of "slipping reconnection" by \citet{Dudik2014}.

The magnetic flux integrated in the brightening pixels, and that in the dimming pixels, is measured, and illustrated in Figure~\ref{fig:0712map}b, showing that, in this event, the onset of flare reconnection starts at around 16~UT, prior to rapid rise of the dimming. The flare reconnection rate peaks at 16:20~UT, and stops around 17~UT, when a total amount of 7$\times 10^{21}$ Mx has been reconnected. The dimming flux is also measured, which rises quickly after 16:20~UT, and the dimming rate peaks around 17~UT, with a total amount of dimming flux reaching 5$\times$10$^{20}$ Mx encompassed in the dimming ribbons. The accompanied CME is observed by STEREO; it exhibits a rapid acceleration starting from 16:10~UT and reaching the maximum at 16:20~UT \citep{Zhu2020}, when the rate of flare reconnection peaks.  

The dimming ribbons are most prominently observed in EUV 304, 171, and 193 passbands. Figure~\ref{fig:0712slit} shows the time-distance diagrams of the normalized brightness produced along 5 slits across each {\em dimming ribbon}; the locations of the slits are indicated by horizontal or vertical bars in Figure~\ref{fig:0712map}a. In particular, the diagrams in the 304~\AA\ passband clearly demonstrate dimming after impulsive brightening. Figure~\ref{fig:0712epoch} shows the epoch plot of the pixel light curves in the two dimming ribbons, or the evolution of the brightness (normalized to the base brightness) with respect to the time of the peak brightness. The figure illustrates the timescales of the impulsive brightening followed by rapid dimming, both within a couple of minutes. In most of these places, dimming does not recover for more than a few hours.

The tempo-spatial sequence of the brightening and dimming in this event suggests the scenario of magnetic reconnection between an erupting structure from the core region and the overlying arcades, or the strapping field. The two dimming ribbons outline the photospheric intersection of separatrices of the complex multi-polar magnetic field where reconnection tends to occur \citep{Dudik2014}. In particular, a potential field extrapolation suggests that the dimming ribbon in the negative magnetic field in the east outlines the feet of an arcade connecting the dimming ribbon with the outer edge of the flare ribbon in positive magnetic field, and the dimming ribbon in positive magnetic field in the west maps the feet of another set of arcades with their conjugate feet in the negative magnetic field north of the flare ribbon (Cooper Downs, private communication). These overlying arcades have to open up for the underlying structure to escape the solar corona.


In summary, in this eruptive event, the dimming analysis does not reveal significant pre-eruption dynamics of coronal magnetic structures. Dimming occurs after the onset of flare reconnection in the core region. The pair of dimming ribbons extended outward from flare ribbons most likely map the feet of several overlying arcades, which open up by reconnecting with the erupting structure. In this course, 5$\times 10^{20}$ Mx overlying flux, which is 10\% of the flare reconnection flux, is removed from the path of the eruption. 

\section{Pre-eruption Dimming Indicative of Gradual Expansion} \label{sec:20110621}

\begin{figure}    
\centerline{\includegraphics[width=1.0\textwidth,clip=]{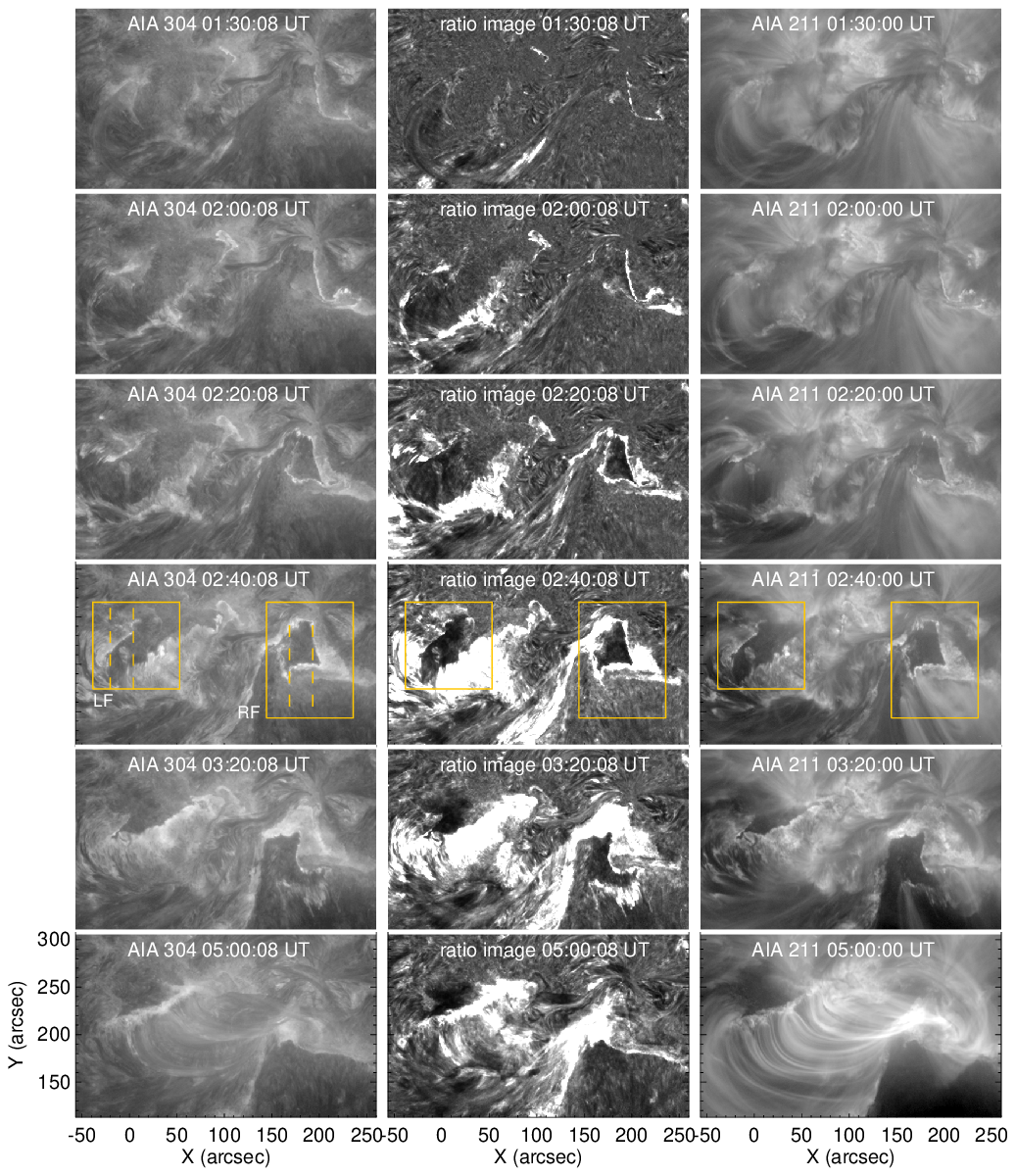}
              }         
	\caption{Overview of the SOL20110621 eruptive flare observed by AIA, in EUV 304~\AA\ passband (left) and 211~\AA\ passband (right). Also shown are base ratio images in the 304~\AA\ passband (middle), with the images normalized to the base image, which is the average of images from 1:00~UT to 1:20 UT. The two orange boxes denote regions of prominent dimming at the left (LF) and right (RF) feet of a filament visible in the EUV images. The four vertical dashed bars in the left panel mark the slits along which the time-distance diagrams are produced in Figure~\ref{fig:0621slit}.}
 \label{fig:0621overview}  
   \end{figure}

\begin{figure}    
\centerline{\includegraphics[width=1.0\textwidth,clip=]{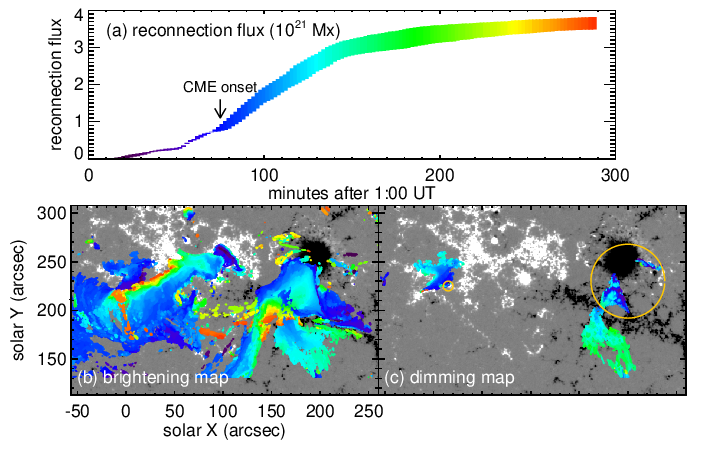}
              }         
	\caption{(a) The magnetic flux, or reconnection flux, measured in brightening pixels. (b)-(c) The time evolution of the brightening and dimming mapped on a photospheric magnetogram of the line-of-sight magnetic field component. At each location, the color indicates the time of the onset of brightening (b) or dimming (c), defined in the text. The time of the color scheme can be read up in panel (a).} 
 \label{fig:0621map}  
   \end{figure}

\begin{figure}    
\centerline{\includegraphics[width=1.0\textwidth,clip=]{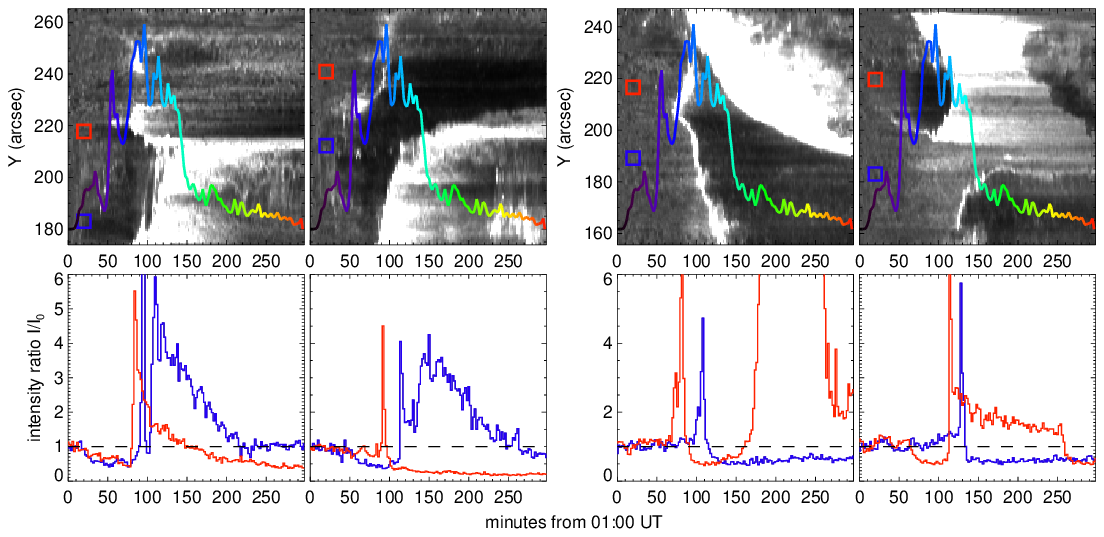}
              }         
	\caption{Top: time-distance diagrams of the normalized brightness along four slits indicated in Figure~\ref{fig:0621overview}, two in the left foot region (left), and two in the right foot region (right), superimposed with the reconnection rate $\dot{\psi}_{rec}$. The color code of $\dot{\psi}_{rec}$ is the same as in Figure~\ref{fig:0621map}. Bottom: light curves of the normalized brightness at a few locations indicated by the red or blue symbols in the top panels.}
 \label{fig:0621slit}  
   \end{figure}

A C-class eruptive flare occurred on 2011 June 21. It has been studied by \citet{Zhu2020, Vievering2023}. The associated CME is best observed by STEREO-A from the limb during its early phase; following the trajectory of the CME, its onset is determined to be at 2:15~UT \citep{Zhu2020}. Figure~\ref{fig:0621overview} shows evolution of the event observed in EUV 304~\AA\ (left) and 211~\AA\ (right) passbands. A filament is visible prior to eruption, and is erupted around the time of the CME onset. At the eruption, two flare ribbons are brightened like depicted in a standard model; the eruption also causes coronal dimming in a large area to the south of the source region. In this study, we focus on the core region where the eruption is originated, with the attempt to understand the early-phase evolution.

This event exhibits a variety of brightening and dimming signatures different from those of the SOL20120712 event. Seen in Figure~\ref{fig:0621overview}, prominent dimming signatures are observed in both passbands, and in particular illustrated in the EUV 304~\AA\ base ratio images (middle panels). Dimming primarily occurs in two regions demarcated by the orange boxes, and are located at the two ends of the filament. The dimming morphology resembles the twin-dimming or core-dimming geometry similar to events reported by \citet{Webb2000, Cheng2016, Wang2019}, suggesting that the twin-dimming regions might map the feet of a flux rope. It is also noted that, along the edge of the dimming cores at the far ends of the two flare ribbons, brightening occurs as early as 1:30~UT, well before the onset of eruption and flare reconnection producing the two ribbons.

In this event, the brightening is relatively weak compared with the X-class flare studied in S\ref{sec:20120712}, therefore we use EUV 304~\AA\ observations, which is more sensitive to weak brightening than 1600~\AA\ passband, to identify brightening signatures at the base of the corona. Two kinds of brightening signatures are observed: flare ribbons underlying closed flare loops exhibit impulsive brightening followed by a prolonged brightening that gradually decays over more than 10 minutes; on the other hand, a number of pixels of weak and brief brightening are also identified. Figure~\ref{fig:0621map}b shows the timing of brightening\footnote{The onset time of the brightening depends on the empirical threshold used to identify brightening, and it is typically within 0-3 min before the time of the peak brightness.} superimposed on a photospheric magnetogram of $B_{los}$. The magnetic flux integrated in the brightening pixels is displayed in Figure~\ref{fig:0621map}a. Note that due to some mixture of brightening of coronal features, especially during the filament eruption between 2:15-2:40~UT, the measured flux is an over estimate of the reconnection flux $\psi_{rec}$ \citep[see ]['s measurement using 1600~\AA\ images, which likely under-estimates the total $\psi_{rec}$]{Zhu2020}. It is noted that, in this event, the initial brightening occurs adjacent to the twin-dimmings at the far ends of the later formed two flare ribbons, suggesting that pre-eruption reconnection might take place between the hypothetic flux rope with ambient fields. 
The amount of the flux estimated from the early brightening signatures prior to eruption (2:15~UT) is about one fifth of the total reconnection flux.


\begin{figure}    
\centerline{\includegraphics[width=1.0\textwidth,clip=]{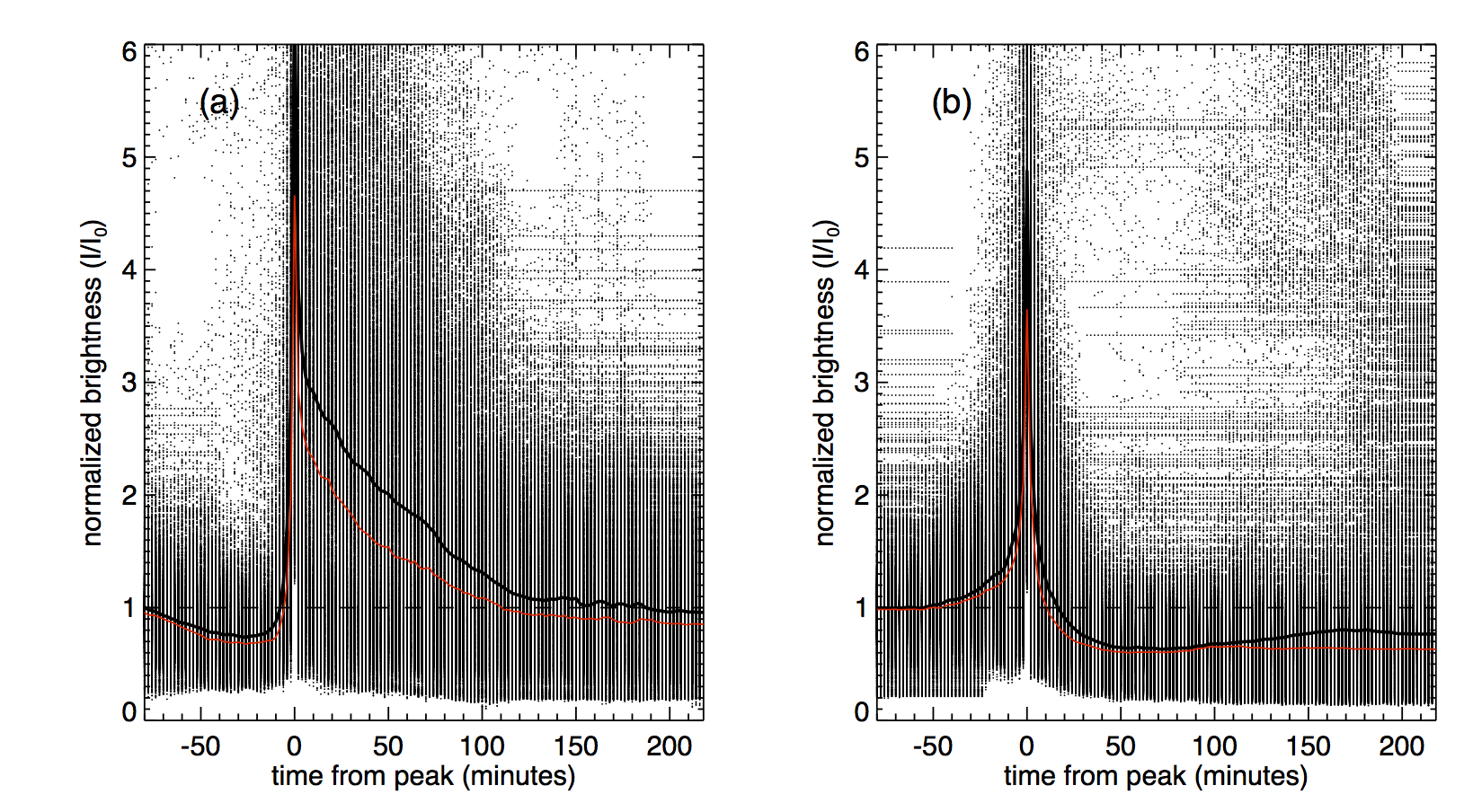}
              }         
	\caption{Epoch plots of dimming light curves in two groups exhibiting the pre-eruption gradual dimming (left), and post-eruption impulsive dimming (right). Solid black and red curves indicate the average and median light curves, respectively, of all light curves in a group.}
 \label{fig:0621epoch}  
   \end{figure}

We identify dimming using EUV 304~\AA\ imaging observations, following the method outlined in S\ref{sec:method}. We form the base image as the average of the images between 1:00 and 1:20~UT, and define dimming as persistent attenuation at $\le 80\%$ of the base brightness $I_0$ at the same location for more than 10 minutes. The onset of dimming, which is illustrated in Figure~\ref{fig:0621map}c, is identified as the time when the brightness starts to fall below 80\% of $I_0$. The analysis reveals two groups of dimming signatures; one group exhibit gradual dimming over tens of minutes prior to the eruption, and in the other group, dimming occurs after the onset of the eruption (at 2:15~UT), often preceded by impulsive brightening. 

Figure~\ref{fig:0621slit} shows the time-distance diagrams of the normalized brightness along a few slits denoted in Figure~\ref{fig:0621overview}, and the light curves of a few selected pixels along the slits exhibiting various dimming-brightening sequences similar to those in Figure~\ref{fig:pxlcv}. The epoch plots of the two groups of dimming light curves are presented in Figure~\ref{fig:0621epoch}, showing the evolution of the pixel brightness with respect to their peak time. The first group exhibits gradual dimming (Figure~\ref{fig:0621epoch}a), with the observed onset of the dimming ranging between 10 - 100 min (average at $54\pm19$ min) before they become brightened. The other group of dimming pixels are characterized by rapid dimming following impulsive brightening (Figure~\ref{fig:0621epoch}b), and the peak dimming depth is about 42\% $\pm$ 12\%. In most places, dimming has not recovered after more than three hours. 

The locations of the two groups of dimmings are marked in Figure~\ref{fig:0621map}c. The pre-eruption gradual dimmings are within the orange circles; they are mostly clustered in the right region in negative magnetic fields and are bounded by early brightening outlining a triangle shaped hook. In the left region in positive magnetic fields, only a small area right next to the flare ribbon exhibits pre-eruption gradual dimming. The post-eruption dimming occupies a larger area in both the left and right feet regions. It appears that, at the onset of eruption, the pre-eruption dimmings spread out into post-eruption dimmings, with rapid brightening during the transition. 

The post-eruption dimming light curves in the second group are similar to those in the SOL20120712 event, suggesting that the erupting structure reconnects with overlying fields and escapes from the corona. However, the geometry of the dimming is different; in this event, the post-eruption dimming appears to map the feet of the erupting structure that has shifted outward after the onset of the flare reconnection, rather than the feet of overlying arcades. This should be confirmed with future data-driven MHD modeling.

The pre-eruption gradual dimming at the feet of the filament are likely signatures of quasi-equilibrium expansion of a coronal structure. We may estimate the mean speed of the expansion from the slope of the dimming depth \citep{Qiu2017b}.
At the base of the corona, the optically-thin transition-region emission may be modeled as the ``pressure gauge" \citep[][and references therein]{Qiu2013}, so that the brightness is proportional to the mean pressure of the overlying corona. As the overlying corona gradually expands, dimming in terms of the ratio of the brightness to the base brightness, $R \equiv I/I_0$, roughly varies as $\dot{R} \approx \alpha (L_0/L)^{\alpha}(\langle v\rangle/L)$, where $\alpha \ge 1 $ is a factor close to unity dependent on the gas expansion model, $L$ is the equivalent height along the line of sight, and $L_0$ being the height before the expansion. For slow (subsonic) expansion, approximating $L \approx L_0$ and $\alpha \approx 1$, the expansion velocity goes roughly $\langle v\rangle \approx \dot{R}L_0$. Fitting the dimming light curves by $R(t) \equiv I(t)/I_0 = R_0 - \dot{R}t$, we derive the dimming slope $\dot{R} \approx 0.001 - 0.01$ min$^{-1}$, the average being 0.003 min$^{-1}$, or about 0.3\% decrease in brightness per minute. For $L_0$ of a few tens of Mm, $\langle v \rangle$ is about a few kilometers per second. The estimated sub-sonic expansion speed is consistent with the expansion speed directly observed from limb observations of a different event which also exhibits persistent gradual dimming, when viewed from the disk, prior to the eruption \citep{Wang2019}. 

In summary, the eruptive event SOL20110621 displays different brightening-dimming geometry and sequence from those of the SOL20120712 event. Both pre-eruption and post-eruption dimming occurs at the feet of a filament adjacent to flare ribbons, and its geometry resembles the core twin-dimming. The twin-dimming likely maps the conjugate feet of a coronal structure that gradually expands and then erupts, and this structure carries the total (axial) flux of up to a few times 10$^{19}$ Mx, estimated by integrating the flux in the dimming area. The early brightening surrounding the gradual dimming cores may indicate the coronal structure expanding through (and interact with) the ambient field.

, 




\section{Summary}{\label{sec:summary}}

In this paper, we present an experiment that analyzes the brightening and dimming signatures in the lower atmosphere, and uses the tempo-spatial sequence of these signatures to identify overlying magnetic structures undergoing dynamic evolution in the early phase of their eruption, such as reconnection or expansion. The experiment is applied to two eruptive events, whose pixel light curves exhibit a variety of brightening-dimming sequences.

For the SOL20120712 event, we do not find signatures indicative of dynamic evolution prior to flare reconnection, which occurs nearly simultaneously with eruption. The erupting structure then reconnects with the overlying arcades, producing impulsive brightening and rapid dimming at the feet of the arcades as the overlying field lines are opened up and removed from its path out. We do not find signatures that likely map the feet of the erupting structure in this event \citep[e.g.][]{Gou2023}. This is possibly due to the stringent criterion in selecting dimming pixels in this study, which requires persistent dimming (for more than 10~min) observed with low-temperature lines characterizing signatures at the base of the corona, such as in the transition region. These requirements are reinforced to help minimize projection effect and noise fluctuations; on the other hand, dimmings at these lines, particularly in the EUV 304~\AA\ passband, are relatively weak. Therefore, dynamic dimming variations on short timescales may not be picked out with the method used in this paper, and the estimated dimming flux is likely the lower-limit.

The SOL20110621 event exhibits pre-eruption gradual dimming at the feet of a filament that is erupted later, and the pre-eruption dimming is accompanied by early brightenings outlining the boundary of the gradual dimming at the far ends of flare ribbons formed later. These are likely signatures of quasi-equilibrium expansion of a coronal structure through the ambient field, which persists for tens of minutes before the explosive loss of equilibrium leading to eruption. If this structure embodies a pre-existing flux rope, the gradual expansion and accompanying reconnection with overlying fields \citep[e.g.][]{Longcope2014b} would bring it to a larger height in favor of the onset of the ideal instability, such as the torus instability \citep{Kliem2014}.

Comprehensive modeling \citep[e.g.][]{Rempel2023} is needed to reconstruct the three-dimensional magnetic configuration and the evolution towards eruption in both events. The identified tempo-spatial sequence of dimmings and brightenings provide additional observational constraints, or the observed boundary conditions, for successful modeling of real eruptive events. 

\begin{acknowledgments}

This work was motivated by discussions during the coronal dimming workshop sponsored by the International Space Science Institute (ISSI) at Bern Switzerland, and during the magnetic flux rope workshop sponsored by the Institute for Space-Earth Environmental (ISEE) Research in Nagoya University of Japan.
J.Q.\ was supported by NASA grants Nos.\ 80NSSC22K0519 and 80NSSC23K0414. {\em SDO} is a mission of NASA's Living With a Star Program.

\end{acknowledgments}



\bibliography{dimming}{}
\bibliographystyle{aasjournal}



\end{document}